\documentclass[prl,twocolumn,superscriptaddress,showpacs,floatfix]{revtex4}

\usepackage{graphicx}
\usepackage{dcolumn}
\usepackage{bm}

\begin{document}

\title{{\it Ab initio} shell-model calculation for $^{18}$O
in a restricted no-core model space}


\author{S. Fujii}
 \email{sfujii@rche.kyushu-u.ac.jp}
\affiliation{
Center for Research and Advancement in Higher Education,
Kyushu University,
Fukuoka 810-8560, Japan
}

\author{B. R. Barrett}
\affiliation{
Department of Physics,
University of Arizona,
Tucson, Arizona 85721, USA
}

\date{\today}

\begin{abstract}
We perform an {\it ab initio} shell-model calculation for $^{18}$O
in a restricted no-core model space, microscopically deriving
a two-body effective interaction and introducing a minimal refinement
of one-body energies in the $spsd$ or $spsdpf$ model space.
Low-lying energy levels, except for the experimental $0_{2}^{+}$ and
$2_{3}^{+}$ states,
are better described in the $spsdpf$ space than in the $spsd$ space.
The structure of low-lying energy levels is discussed with an emphasis
on many-particle many-hole states beyond the four-particle two-hole
configuration.
\end{abstract}


\maketitle

The nuclear shell model is a successful method to describe
the structure of the nucleus~\cite{Mayer49,Jensen49}.
Recently, the understanding of the nuclear shell structure has been
increased by elucidating complicated properties,
especially the tensor force,
of underlying nuclear interaction~\cite{Otsuka05}.
On the other hand, as the numerical calculation becomes larger
and more microscopic,
the meaning of the word ``shell" has become vague.
One of the {\it ab initio} methods to describe nuclei starting with
the bare nucleon-nucleon (NN) force in free space is
the no-core shell model (NCSM)~\cite{Navratil00,Navratil02}.
In the sense of using harmonic-oscillator (h.o.) Slater-determinant
basis states, the NCSM may be regarded as a shell model.
However, its basis states span a huge model space,
which is taken as large as possible,
until the result of the diagonalization converges.
Therefore, the meaning of the shell in the NCSM is somewhat different
from the usual one in the conventional shell model.

While the formalism of the NCSM for heavier nuclei has been developed,
its full application has been limited to nuclei around the mass
number $A=12$ ($p$-shell nuclei), due to an explosive expansion of
the Hamiltonian matrix to be diagonalized.
If one explores nuclei heavier than $A=12$, some approximations or
truncations must be employed and the results obtained are no longer
exact.
However, those results may still contain some important physics.
It is well-known that, even in a much more restricted model space,
such as the $sd$ or $pf$ space, the conventional shell-model study
with an optimal effective interaction can explain many experimental
data~\cite{Poves81,Brown88, Hjorth-Jensen95, Honma04}.
Thus, it is our purpose in this study to bridge the gap
between the two approaches.

Recently, some attempts have been proposed along this line.
One is a shell-model calculation in the no-core $spsdpf$ space
combined with a minimal refinement of the one-body
energies~\cite{Fujii07}.
This no-core type of shell model has been applied to the neutron-rich
carbon isotopes around $^{16}$C, and it has been shown that low-lying
energy levels are well described.
Another is an {\it ab initio} shell model with a core in which
a $p$-shell effective Hamiltonian for light nuclei is constructed
utilizing a NCSM result in a sufficiently large model
space~\cite{Lisetskiy08}.
In both methods, no experimental information about the energy levels
is used, and a double unitary transformation is performed to
microscopically derive the effective interaction.
In this Letter, we employ the former method and investigate the
structure of $^{18}$O.

Since  $^{18}$O is a typical $sd$-shell nucleus, which has
the structure of two neutrons and a core nucleus, there have been
many theoretical studies.
As for the shell model with a microscopic effective interaction,
the work of Kuo and Brown is the pioneering study~\cite{Kuo65}.
They constructed the effective interaction for the $sd$ model space
and showed that low-lying energy levels are well described.
However, even in this early stage, 
the importance of including a larger space for the intermediate states
in determining the effective interaction
was pointed out by others~\cite{Barrett70,Vary73}.

In Ref.~\cite{Coraggio09}, a recent $sd$ shell-model calculation for
$^{18}$O using a modern NN interaction with the folded diagram theory
has been reported.
Although this method seems to be successful in describing low-lying
energy levels,
intruder states having many-particle many-hole dominant configurations
may not be properly reproduced due to the limited $sd$ model space.
If one wants to describe such many-particle many-hole states
explicitly, several major shells must be included in the model space.
In this sense, our present work in a {\it restricted} no-core model
space would be preferable for the simultaneous description of
the ground and complicated excited states.

Here, we outline the method of deriving the effective interaction in
our restricted no-core space. The details are given in
Refs.~\cite{Fujii04,Fujii07}.
First we derive the two-body effective interaction from a realistic NN
interaction in a large model space $P_{1}$,
solving the decoupling equation~\cite{Suzuki80}
between the model space $P_{1}$ and its compliment $Q_{1}$.
The $P_{1}$ space consisting of the two-body states is defined by
a boundary number $\rho _{1} \leq 2n_{1}+l_{1}+2n_{2}+l_{2}$
with the h.o. quantum numbers \{$n_{1},l_{1}$\} and \{$n_{2},l_{2}$\}
of the two-body states.
Note that the full space of the two-body states $P_{1}+Q_{1}$,
which is necessary to accurately take into account the strong repulsion
effect of the NN force at short distance reaches about
300 $\hbar \Omega$ in h.o. energy.
Therefore, in general, the diagonalization of the original many-nucleon
Hamiltonian in such an energy range could not be performed
in the foreseeable future.
The effective interaction determined in this first step corresponds to
a well-behaved interaction that renormalizes the effect of
the short-range repulsion.

If one uses this effective interaction for a sufficiently large model
space in the diagonalization of the Hamiltonian,
this leads to the NCSM.
However, this type of approach cannot be applied to heavier nuclei,
since the dimension of the matrix becomes huge.
Therefore, in order to construct an effective interaction suitable for
a no-core type of shell-model calculation in a tractable model space,
we again solve the decoupling equation between a smaller model space
$P_{2}$, such as the $spsd$ or $spsdpf$ space, and its compliment
$Q_{2}$ using the two-body effective interaction determined
in the first step, where $P_{1}=P_{2}+Q_{2}$.
Note that the boundary number $\rho _{1}$ is taken to be $\rho _{1}=6$
in the present work and, thus, is not so big,
although this number should be taken as large as possible
in the usual sense of the NCSM.
The smaller model space $P_{2}$ and its compliment $Q_{2}$ are not
separated by energy but by state, and, thus, some two-body states
in the model space are degenerate with those in its complement.
This degeneracy often causes a difficulty in determining
the effective interaction.
Furthermore, in contrast to the usual NCSM, we solve the decoupling
equation by introducing a self-consistent one-body potential for
$^{16}$O and use it to construct a suitable effective interaction
in the small model space for nuclei around $^{16}$O.
This self-consistent procedure also makes the problem more difficult
to solve, if the degeneracy is big.
Thus, in this second step of the calculation, the self-consistent
potential is introduced only for the small model space $P_{2}$.
The self-consistent potential is also calculated for
the large model space $P_{1}$ in the first-step decoupling.
Although we decided to take the {\it small} number $\rho _{1}=6$
to make the degeneracy as small as possible, this artificial
truncation is compensated to some extent by introducing
a minimal refinement of one-body energies as explained later.

Consequently, our shell-model Hamiltonian is composed of the sum
of the one-body kinetic energy
(the minimal refinement of one-body energies),
the two-body effective interaction for the
$P_{2}$ space, and the center-of-mass (c.m.) Hamiltonian
which is multiplied by a large value $\beta _{\rm c.m.}$ to separate
the spurious c.m. motion from the physical states~\cite{Gloeckner74}.

\begin{figure}[t]
\includegraphics[width=.290\textheight]{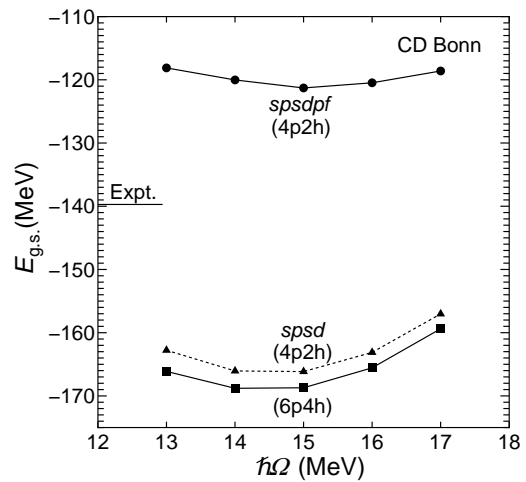}
\caption{\label{fig:o18egs} The $\hbar \Omega$ dependence
of the calculated ground-state energies of $^{18}$O.}
\end{figure}

In this work, we use the CD-Bonn potential~\cite{Machleidt96}
as the original NN interaction.
In Fig.~\ref{fig:o18egs},
we present the dependence of the ground-state energy of $^{18}$O
on the h.o. frequency $\hbar \Omega$.
The effect of the minimal refinement has not yet been
included in the results.
The results for three different model spaces used
in the diagonalization are shown: up to
the four-particle two-hole (4p2h) configuration in the $spsdpf$ space,
the 4p2h $spsd$, and the 6p4h $spsd$ from the unperturbed ground state
of $^{16}$O.
The result at the energy minimum point for the $spsdpf$ space shows
a less attractive energy by about $20$ MeV than the experimental value,
while the results for the $spsd$ space are much more attractive.
It is known that the CD-Bonn potential underbinds nuclei
in high-precision calculations for few-nucleon systems.
Our result for the $spsdpf$ space is consistent with this tendency.
As expected, the $spsdpf$ result is more favorable
than the $spsd$ results.
In the $spsd$ calculations, the overbinding of the energy could be
compensated by introducing a repulsive and sizable three-body
effective interaction, but such a calculation cannot be easily done.
For all three types of calculation,
there appear energy minima at around $\hbar \Omega =15$ MeV.
Therefore, we choose this value as the optimal one
for all the results in this study.

The ground-state energy for the 4p2h $spsdpf$ case at the optimal value
of $\hbar \Omega$ is about $-120$ MeV.
We note, however, that this number has an ambiguity
due to several effects.
In addition to some approximations and truncations in deriving
the effective interaction and performing the diagonalization,
there is a problem regarding the removal of the spurious c.m. motion.
Unlike the usual NCSM approach, our shell-model space is not given
by energy but by state.
In such a state separation, the influence of the spurious c.m. states
on the energy levels of intrinsic states cannot be completely removed
by a suitable value of $\beta _{\rm c.m.}$.
This problem has recently been examined in NCSM and
coupled-cluster calculations by Roth {\it et al.},
and they have shown that the c.m. effect is non-negligible
in restricted and incomplete energy spaces~\cite{Roth08}.
However, if we look at the relative structure from the ground state,
this effect is expected to be small for a sufficiently large value of
$\beta _{\rm c.m.}$, as in the study of carbon isotopes~\cite{Fujii07}.

\begin{figure}[t]
\includegraphics[width=.290\textheight]{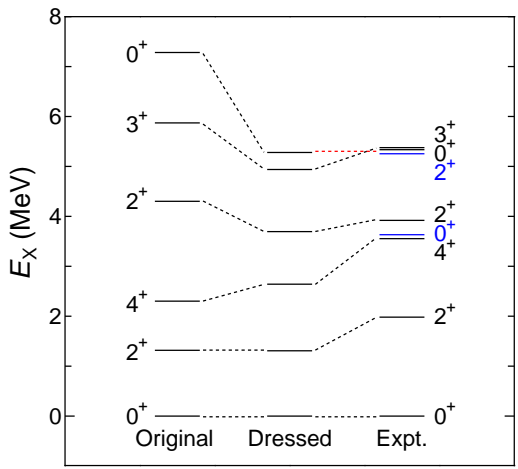}
\caption{\label{fig:o18levels} Low-lying energy levels of the
positive-parity states in $^{18}$O.
}
\end{figure}

In the calculation of the energy levels of $^{18}$O,
it is important to investigate the neutron single-particle energies
for the lowest three states in $^{17}$O.
If the calculated single-particle levels in $^{17}$O
are not comparable to the experimental values,
the calculation for $^{18}$O would be questionable.
We have found that, in the present shell-model space,
the calculated single-particle levels are not so satisfactory.
On the other hand, we know that,
if we compute the single-particle energies using the CD-Bonn potential
within the framework of the unitary-model-operator approach (UMOA),
the results are close to the experiments~\cite{Fujii04}.
The UMOA can be regarded as a sort of e$^{S}$ method or
coupled-cluster theory of the Hermitian type~\cite{Suzuki94}.
In the UMOA, the calculation of the single-particle energy
is feasible in a sufficiently large model space,
although it is hard to obtain complicated states.
Therefore, in order to include the effect of the correlation energy of
the large space into the present restricted one in a simple way,
we change the one-body matrix elements of the neutron
in the $sd$ shell, utilizing single-particle information obtained
by the UMOA.
This variation is the ``minimal refinement".

The magnitude of the minimal refinement is determined so as to
reproduce the binding energies of the lowest
single-particle $5/2^{+}$, $1/2^{+}$, and $3/2^{+}$ states in $^{17}$O
obtained within the UMOA, as a consequence of the shell-model
diagonalization in the present space.
Denoting the one-body energy for a single-particle orbit $j$ of
the neutron as $\epsilon ^{n}(j)$,
we need these variations for the 4p2h $spsdpf$ case to be
$\Delta \epsilon ^{n}(0d_{5/2})= 0.99$,
$\Delta \epsilon ^{n}(1s_{1/2})=-0.42$, and
$\Delta \epsilon ^{n}(0d_{3/2})=-1.04$ MeV.
These are added to the original $\epsilon ^{n}({j})$'s, namely,
the h.o. kinetic energies.
We have checked that the results for the excited state energies above
the ground state are hardly changed, even if we introduce similar
refinements into the proton orbits of the $sd$ shell and
the hole orbits for the neutron and the proton.

In Fig.~\ref{fig:o18levels}, we illustrate calculated
low-lying energy spectra of the positive-parity states
for the 4p2h $spsdpf$ case, together with the experimental levels.
The energy levels in the first and second columns, labeled by
``Original" and ``Dressed", denote the results without and with the
minimal refinement, respectively.
For the case ``Original", the excited levels above $E_{X}=3$ MeV lie
higher than the corresponding experimental ones, while those below
$E_{X}=3$ MeV lower than the experiments.
By introducing the minimal refinement, the results on the whole
become closer to experiment.
A similar tendency was seen in the calculation for the neutron-rich
carbon isotopes~\cite{Fujii07}.

\begin{table}[t]
\caption{\label{tab:o18levels} The excitation energies
of low-lying positive-parity states in $^{18}$O.
The calculated results are for the cases (a) 4p2h $spsd$,
(b) 6p4h $spsd$, and (c) 4p2h $spsdpf$.
All energies are in MeV.}
\begin{ruledtabular}
    \begin{tabular}{ccccc}
       $J^{\pi}$ &   (a)  &   (b)  &   (c)  &  Expt. \\ \hline
       $3^{+}$   & $3.68$ & $3.84$ & $4.94$ & $5.38$ \\
       $0^{+}$   & $4.13$ & $4.36$ & $5.28$ & $5.34$ \\
       $2^{+}$   & $$     & $$     & $$     & $5.25$ \\
       $2^{+}$   & $3.00$ & $3.09$ & $3.70$ & $3.92$ \\
       $0^{+}$   & $    $ & $$     & $$     & $3.63$ \\
       $4^{+}$   & $1.48$ & $1.57$ & $2.64$ & $3.55$ \\
       $2^{+}$   & $0.67$ & $0.74$ & $1.31$ & $1.98$ \\
    \end{tabular}
\end{ruledtabular}
\end{table}

To demonstrate the advantage of the calculation in the
$spsdpf$ space over the $spsd$, as in Fig.~\ref{fig:o18egs},
the results of the low-lying energy levels for the three spaces
in the ``Dressed" case are given in Table~\ref{tab:o18levels}.
Note that the magnitudes of the minimal refinement of the one-body
energies depend on the model-space size employed.
The magnitudes for the 4p2h $spsdpf$ have been given before;
those for the 4p2h $spsd$ are
$\Delta \epsilon ^{n}(0d_{5/2})= 2.12$,
$\Delta \epsilon ^{n}(1s_{1/2})=-3.14$, and
$\Delta \epsilon ^{n}(0d_{3/2})=-1.81$ MeV,
and for the 6p4h $spsd$,
$\Delta \epsilon ^{n}(0d_{5/2})= 2.42$,
$\Delta \epsilon ^{n}(1s_{1/2})=-2.59$, and
$\Delta \epsilon ^{n}(0d_{3/2})=-1.30$ MeV.
These values are determined in each restricted model space
so as to reproduce the UMOA results.
The energy spacings for the 4p2h $spsd$ case are
rather compressed compared to the experimental data.
The inclusion of the $6p4h$ configuration does not cause
any significant change in the relative spacings.
The results for the $0_{2}^{+}$ state for the $spsd$ cases
are fairly close to the experimental $0_{2}^{+}$
rather than the experimental $0_{3}^{+}$.
However, by including the $pf$ shell,
the calculated $0_{2}^{+}$ state becomes close to
the experimental $0_{3}^{+}$, and the energy spacings are expanded.
Then the results become, on the whole,
close to the experimental values.
This suggests the risk of using the experimental single-particle
energies in the conventional shell model,
leading to an accidental reproduction of energy levels of
the wrong structure~\cite{Hofmann74}.

As shown before, the minimal refinements for the $spsd$ cases
are considerably larger than the $spsdpf$ ones.
This means that the minimal refinements in the $spsd$ cases
are not so realistic.
In general, the effect of the many-body effective interaction becomes
bigger as the size of the model space becomes smaller.
Therefore, in the $spsd$ space, the change of only
the one-body energy may be too simple a treatment.
In such a small model space, we should change not only
the one-body energy but also the two-or-more-body matrix element.

\begin{table}[t]
\caption{\label{tab:o18_2nd0+} The excitation energies $E_{\rm x}$
in MeV and the probabilities of the 4p2h configuration $P_{\rm 4p2h}$
for the lowest 4p2h-dominated $0^{+}$ state in $^{18}$O,
for the configurations up to 4p2h, 6p4h, and 8p6h in the $spsd$ space.}
\begin{ruledtabular}
    \begin{tabular}{cccc}
      Configuration ($spsd$) & 4p2h     &   6p4h   & 8p6h  \\ \hline
      $E_{\rm x}$    & $41.75$ & $31.50$ & $30.29$ \\
      $P_{\rm 4p2h}$ & $0.996$ & $0.855$ & $0.816$ \\
    \end{tabular}
\end{ruledtabular}
\end{table}

Interestingly, in the $spsdpf$ case, the result of the $0_{2}^{+}$
state appears at near the experimental $0_{3}^{+}$,
and there is no calculated $0^{+}$ state near the experimental
$0_{2}^{+}$.
Furthermore, we cannot find a $2^{+}$ state near the experimental
$2_{3}^{+}$.
These experimental $0_{2}^{+}$ and $2_{3}^{+}$ levels have been treated
as (deformed) 4p2h states in shell-model calculations~\cite{Hofmann74},
while these states have been considered to be well-developed
$^{14}$C$+\alpha$ cluster states in Refs.~\cite{Alhassid82,Gai83}.
In our calculation, the 4p2h configurations are included.
Nevertheless, there occurs no indication
that our 4p2h states come down to the low-lying energy region.

There may be some reasons for the disappearance.
One is the simple treatment of the minimal refinement,
even in the $spsdpf$ space.
Another is the effect of higher configurations, such as the 6p4h,
on the 4p2h dominant state.
To investigate this effect, we have calculated the
lowest $4p2h$-dominated state in the {\it spsd} case.
In Table~\ref{tab:o18_2nd0+}, we tabulate the excitation energies
and the probabilities of the 4p2h configuration
for the lowest $0^{+}$ state dominated by the 4p2h configuration,
for the configurations up to 4p2h, 6p4h, and 8p6h in the $spsd$ space.
The probabilities of the 4p2h configuration contain the contributions
from all the 4p2h states in each space.
The calculated 4p2h state is the fourth $0^{+}$ state
in the $spsd$ case.
The third $0^{+}$, which is dominantly composed of the $(0d_{3/2})^{2}$
configuration of the neutron,
lies at about $16$ MeV in excitation energy for the three cases.
Although the excitation energies in Table~\ref{tab:o18_2nd0+}
are far from the experimental $0_{2}^{+}$,
a considerable contribution of about $-10$ MeV from the 6p4h
configuration is brought to the excitation energy,
while that from the 8p6h configuration is rather small.
This indicates the importance of also including the 6p4h
in larger spaces, even if the $4p2h$ is the dominant configuration,
when (spherical) h.o. wave functions are used as the basis states.
Such a calculation is a challenging problem to reveal the real
structure of the experimental $0_{2}^{+}$ and $2_{3}^{+}$ states.

In conclusion, low-lying states in $^{18}$O are well described
by our {\it ab initio} shell-model calculation with minimal refinement
in the no-core $spsdpf$ space.
However, a larger model space is necessary to properly describe
the experimental $0_{2}^{+}$ and $2_{3}^{+}$ states.

\begin{acknowledgments}

We thank the Institute for Nuclear Theory, University of Washington,
where this work was initiated, and GSI-Darmstadt for their hospitality.
The numerical calculation was partially carried out on the computer
system at YITP in Kyoto University.
S. F. acknowledges support by a Grant-in-Aid for Young Scientists (B)
(Grant No. 18740133) from JSPS and the JSPS Core-to-Core Program EFES,
and B. R. B., by NSF (Grant No. PHY0555396) and the JUSTIPEN.
\end{acknowledgments}

\end{document}